\documentclass[twocolumn,superscriptaddress,10pt,preprint]{revtex4}
\usepackage{graphicx}
\begin{document}
\title{Bubble spreading during the boiling crisis: modelling and experimenting in microgravity}
\author{V. Nikolayev}\thanks{email: vadim.nikolayev@cea.fr phone: +33 1 40 79 58 26\\ fax: +33 1 40 79 45 23}\author{D. Beysens}
\affiliation{ESEME, Service des Basses Temp\'eratures, DRFMC/DSM/CEA-Grenoble,\\
17 rue des Martyrs, 38054 Grenoble Cedex 9, France} \affiliation{CEA-ESEME, ESPCI-PMMH, 10, rue Vauquelin, 75231
Paris Cedex 5, France}
\author{Y. Garrabos}\author{C. Lecoutre}
\affiliation{CNRS-ESEME, Institut de Chimie de la Mati\`{e}re Condens\'{e}e de Bordeaux,\\
Universit\'{e} de Bordeaux I, Avenue du Dr. Schweitzer, 33608 Pessac Cedex, France}
\author{D. Chatain}\affiliation{ESEME, Service des Basses Temp\'eratures, DRFMC/DSM/CEA-Grenoble,\\
17 rue des Martyrs, 38054 Grenoble Cedex 9, France} \preprint{\small To be published in Microgravity Science and
Technology} \preprint{\small Presented at ELGRA Symposium at Santorini, Greece 21-23/09/2005}
\date{\today}
\begin{abstract}
Boiling is a very efficient way to transfer heat from a heater to the liquid carrier. We discuss the boiling
crisis, a transition between two regimes of boiling: nucleate and film boiling. The boiling crisis results in a
sharp decrease in the heat transfer rate, which can cause a major accident in industrial heat exchangers. In
this communication, we present a physical model of the boiling crisis based on the vapor recoil effect. Under
the action of the vapor recoil the gas bubbles begin to spread over the heater thus forming a germ for the vapor
film. The vapor recoil force not only causes its spreading, it also creates a strong adhesion to the heater that
prevents the bubble departure, thus favoring the further spreading. Near the liquid-gas critical point, the
bubble growth is very slow and allows the kinetics of the bubble spreading to be observed. Since the surface
tension is very small in this regime, only microgravity conditions can preserve a convex bubble shape. In the
experiments both in the Mir space station and in the magnetic levitation facility, we directly observed an
increase of the apparent contact angle and spreading of the dry spot under the bubble. Numerical simulations of
the thermally controlled bubble growth show this vapor recoil effect too thus confirming our model of the
boiling crisis.\end{abstract}\maketitle

\section{Introduction}

During the nucleate boiling, separate gas bubbles form at the heater surface. The liquid phase of the fluid is
thus in the direct contact with the heater and the heat transfer rate is very large. This feature explains the
high technological importance of boiling in the applications where the high rate cooling of the heater surface
is required. Besides the nucleate boiling, another regime of boiling exists: film boiling. During the film
boiling, the heater is covered by a quasi-continuous gas film. This film thermally insulates the heater from the
liquid and the heat transfer rate is much smaller than in nucleate boiling regime. The transition from nucleate
to film boiling thus causes a sharp loss of the heat transfer and the heater can melt down if the heating power
is not controlled. This transition is called ``Boiling Crisis" (BC), or ``CHF phenomenon" since it occurs when
the heat flux $q$ from the heater attains a threshold value $q_{CHF}$ called  ``Critical Heat Flux" (CHF)
\cite{Tong,Straub}. In spite of the considerable amount of studies devoted to this important phenomenon, its
mechanism is not well understood. This difficulty originates from the violence of the fluid motion during the
conventional boiling that on the one hand conceals the mechanisms of bubble growth from detailed observation,
and on the other hand hugely complicates direct numerical simulations. Most still unanswered questions concern
close vicinity of the heating surface, down to the scale of the foot of a bubble growing at the surface of the
heater. The recent revival of the interest to the microgravity boiling in general and to BC in particular is
related to the development of the engine \cite{Vinci} of the upper stage of Ariane 5 launcher that will need to
be restarted under microgravity conditions.

The boiling can be very different since it is dependent on many parameters like system pressure, presence of the
externally imposed fluid flow, temperature in the fluid bulk that can be at saturation or below, thermophysical
properties of the fluid and the heater etc. In all cases, BC begins in a very thin fluid layer (adjacent to the
heater), the thickness of which is small in comparison with the bubble size. The phenomena at this scale ``feel"
only the averaged values of macroscopic flow variables and thus do not depend on details of the flow regime
which should not thus change the BC mechanism. However the contribution of the microscale hydrodynamic motion
(that can change in magnitude with the system pressure, see below) to the BC triggering is still unclear.

For the reasons of industrial importance, boiling is mostly studied at low pressures comparing to the critical
pressure $p_c$ (the pressure of the gas-liquid critical point) of the fluid under study, e.g. for water or freon
at atmospheric pressure. $q_{CHF}$ is then large (of the order of several MW/m$^2$ for water) and the boiling
close to it is indeed extremely violent. However, it is well known \cite{Tong} that $q_{CHF}$ decreases at high
pressures where BC can thus be observed at a smaller heat flux. In addition, the thermal diffusivity is smaller
in this regime, the bubble growth slows down, and the optical distortions disappear due to the slowness of the
fluid motion. From the point of view of the modelling, this slowness permits to simplify the problem by
neglecting the hydrodynamic stresses at the bubble interface, the shape of which can then be calculated in the
quasi-static approximation.

\section{Force of vapor recoil and bubble spreading}

Several groups \cite{Pav,Avk,Stein} proposed the vapor recoil instability \cite{Palmer} as a reason for BC.
Although they did not make clear why and how this instability can induce the spreading of a dry spot, these
authors show that the vapor recoil can be an important factor.

Every fluid molecule evaporated from the liquid interface causes a recoil force analogous to that created by the
gas emitted by a rocket engine. It pushes the interface towards the liquid side in the normal direction. This
force appears because the fluid necessarily expands while transforming from liquid to gas phase. Obviously, the
stronger the evaporation rate $\eta$ (mass per time and interface area), the larger the vapor recoil force. The
vapor recoil force per unit interface area reads \cite{Palmer} $P_r=\eta^2(\rho_V^{-1}-\rho_L^{-1})$, where
$\rho_L$($\rho_V$) signifies liquid (gas) density.

\begin{figure}[htb]
\centering
\includegraphics[width=6cm]{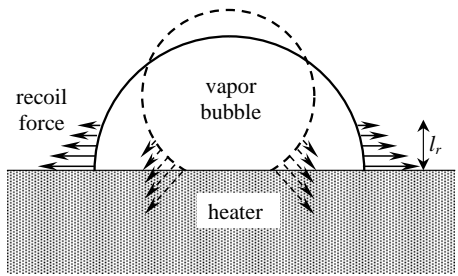}
\caption{Sketch illustrating the vapor recoil effect. The amplitude and direction of the vapor recoil force are
shown by arrows. The thickness $l_r$ of the belt on the bubble surface, in which the vapor recoil force is
important, is exaggerated with respect to the bubble size.} \label{brec}
\end{figure}
Let us now consider a growing vapor bubble attached to the heater surface (Fig.~\ref{brec}). While the
temperature of the vapor-liquid interface is constant and equal to the saturation temperature for the given
system pressure for the pure fluid (see the discussion in \cite{Straub,PRE01,Tadrist} and references therein), a
strong temperature gradient forms near the heating surface. The liquid is overheated in a thermal boundary
layer, and the heat flux $q_L$ at the bubble surface is thus elevated in a ``belt" of the bubble surface
adjacent to the bubble foot. As a matter of fact, most of the evaporation into the vapor bubble is produced in
this belt, whose thickness $l_r$ is much smaller than the bubble radius. Since $\eta\approx q_L/H$, where $H$ is
the latent heat, the vapor recoil near the contact line is much larger than at the other part of the bubble
surface. Consequently, the bubble should deform as if the triple contact line were pulled apart from the bubble
center as shown in Fig.~\ref{brec}. Therefore, under the action of the vapor recoil the dry spot under the vapor
bubble should spread so that the bubble will cover the heater surface.

\begin{figure}[htb] \centering
\includegraphics[width=8.5cm]{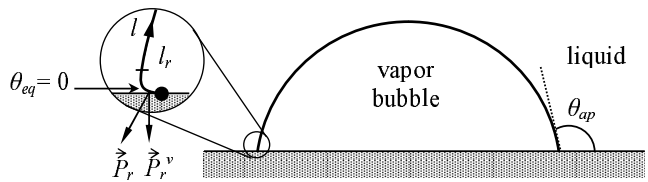}
\caption{Apparent $(\theta_{ap})$ and actual $(\theta_{eq})$ contact angles. The solid circle indicates the
contact line, the zero point for the curvilinear coordinate $l$ measured along the bubble contour. The arrows
show the directions of the vapor recoil force in the vicinity of the contact line and its vertical component
that creates an additional adhesion of the bubble to the heater.} \label{appangle}
\end{figure}

A numerical simulation \cite{IJHMT01} of such a process requires solving of coupled thermal conduction (both in
solid and liquid) and capillary problems. The capillary problem includes a solution for the bubble shape in the
quasi-static approximation in which the local bubble curvature $K$ is defined by the modified Laplace equation
$K\sigma=\Delta p +P_r$. The bubble shape can be determined from the function $K=K(l)$ (Fig.~\ref{appangle}).
$\Delta p$ is the constant difference of the pressures in the bubble and in the rest of the fluid and $\sigma$
is the surface tension. $P_r$ and then $K$ vary along the interface and attain their maxima near the line of the
triple contact liquid-gas-solid. Since curvature is the first derivative of the angle formed by the tangent to
the bubble interface and the horizon, the large $K$ means that this angle varies strongly near the contact line.
The apparent contact angle $\theta_{ap}$ measured at some distance from the contact line can thus deviate
strongly from its actual value $\theta_{eq}$ imposed at the contact line, see Fig.~\ref{appangle}.

The time of residence of a vapor bubble on the heater is an important parameter since the bubble might simply
depart from the heater before having time to spread. The determination of the residence time requires the forces
acting on the bubble to be analyzed. Such an analysis \cite{MI04} shows that the vapor recoil creates an
additional adhesion that prevents the bubble departure as soon as the bubble spreading begins, i.e. when the
vapor recoil becomes important. This adhesion force can be obtained by integrating the vertical component
$P_r^v$ of $\vec{P}_r$ (Fig.~\ref{appangle}) over the bubble interface.

The thermal problem is extremely delicate since most of heat transfer occurs at a scale $l_r$ much smaller than
the bubble size. Extremely fine meshing is thus needed to obtain the correct values of the fluxes near the
contact line. Another difficulty appears since the position of one of the solution domain boundaries, that is of
the bubble interface, is not known in advance and depends (trough $P_r$) on the solution of the thermal problem
itself. The solution of the full problem thus needs to be found by time-consuming iterations.
\begin{figure}[htb]
\centering
\includegraphics[width=7cm]{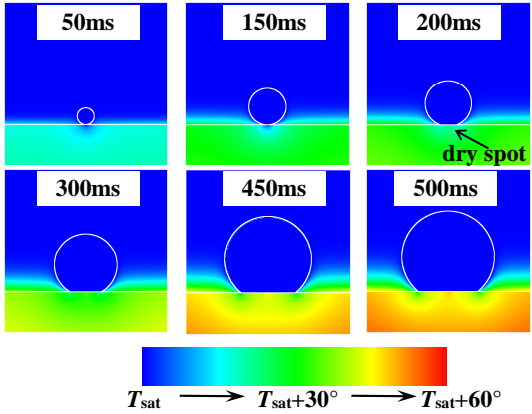}
\caption{Vapor bubble spreading under the influence of vapor recoil simulated for water at 10 MPa. The color
indicates the local temperature with respect to the saturation temperature $T_{sat}=311^\circ$C;
$q=0.1$~MW/m$^2$.} \label{simul}
\end{figure}

The simulated bubble growth is presented in Fig.~\ref{simul}. It can be seen that the dry spot is initially very
small and remains so during the initial growth stage. At about 180~ms the dry spot begins to grow suddenly, i.e.
the bubble spreads. Such a spreading represents the beginning of formation of the gas film characteristic for
BC. The apparent contact angle grows sharply while $\theta_{eq}$ is maintained to be zero throughout the
simulation. Zero value was chosen because the wetting case is most common. The effect is even stronger for
larger contact angles where the dry spot exists from the very beginning. Fig.~\ref{simul} also shows the
formation of a hot spot at the heater surface in the middle of the dry area. This temperature rise illustrates
the already discussed blocking of the heat transfer by vapor.

\section{Experiments in reduced gravity}

To overcome the experimental difficulties encountered during the observations of BC at low pressures, we carry
out our experiments at very high pressures, in the vicinity of the critical point. We take advantage of the
so-called ``critical slowing down" observed near the critical point. In fact, due to the smallness of the
thermal diffusivity, the growing process of a single vapor bubble could be observed during tens of minutes thus
allowing for a very detailed analysis. The CHF value is also vanishing at the critical point \cite{Tong} so that
BC can be examined at a small heat flux $q$ that does not necessarily induce a strong fluid motion which might
hinder the optical observations. However, near-critical bubble growth experiments have an important drawback.
Since the surface tension becomes very low near the critical point, gravity completely flattens the liquid
interface. Weightlessness conditions are thus necessary to preserve the usual convex bubble shape.

In our already performed experiments, the cells are closed and only pure fluid is present in them so that its
total mass and volume remains constant. Unlike the conventional boiling experiments, the gas bubble is not
nucleated but exists already before the cell heating begins. The bubble growth is then observed.

Another particularity of the near-critical systems consists in the symmetry of the co-existence curve
(temperature dependence of $\rho_L$ and $\rho_V$) with respect to the critical density $\rho_c$:
$(\rho_L+\rho_V)/2=\rho_c$. When the average cell density is equal to $\rho_c$, the gas volume remains constant
and equal to one half of the cell volume throughout the heating. The gas mass however increases (and the liquid
mass decreases) during the heating because of the density change. This makes the optical observations even more
convenient.
\begin{figure}[htb]
\centering
\includegraphics[width=8cm]{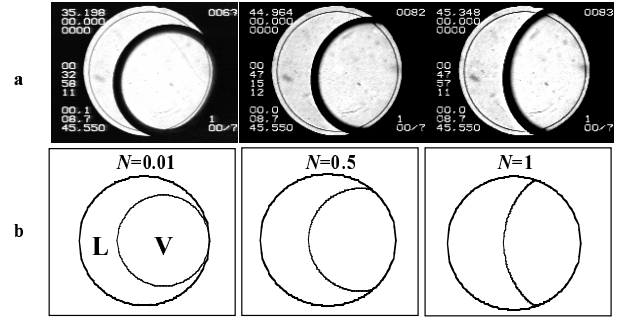}
\caption{Snapshots of the vapor bubble (V) growing in near critical liquid (L), both experimental (a) and
calculated (b) for the given values of the vapor recoil strength $N$. The zero value of the actual contact angle
is imposed while the apparent contact angle attains a large value.} \label{cell_exp-theor}
\end{figure}

Two series of experiments were carried out up to now. The first \cite{PRE01} used SF$_6$ fluid on board of the
Mir space station in the ALICE-2 apparatus designed by the French CNES agency. The choice of SF$_6$ is made for
practical reasons: the critical point of this fluid is $T_c=45.6^\circ$C, $p_c=38$~bar and requires much less
severe conditions for the experiment than for example water ($374^\circ$C, 220~bar). Some of the results of this
experiment are presented in Fig.~\ref{cell_exp-theor}a. The sequential photos of the growing vapor bubble are
taken through the transparent bases of the cylindrical cell, the lateral copper walls of which are being heated.
Spreading of the dry spot under the bubble similar to that in Fig.~\ref{simul} can be seen.

The bubble shapes calculated for different values of the vapor recoil strength $N=\int P_r(l)\mathrm{d}l/\sigma$
(where the integration is performed along the bubble contour) are presented for comparison in
Fig.~\ref{cell_exp-theor}b. Only the capillary problem was solved using $\theta_{eq}=0$ as a boundary condition.
The $P_r(l)$ expression used was that of \cite{EuLet99}.

\begin{figure}[htb]
\centering
\includegraphics[width=8cm]{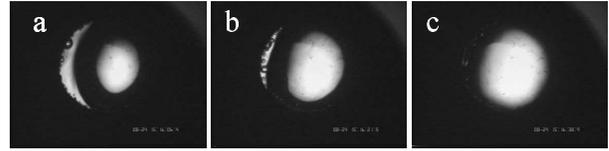}
\caption{Gas spreading at 32K (well beyond the critical region) as visualized through the transparent heater in
magnetic levitation experiment. (a) Beginning of the dry spot growth. (b) Bubble partially spread. (c) Complete
drying of the heater. Nucleated small bubbles (boiling) are visible in (a-b). The volume of the large bubble
remains nearly constant throughout the evolution.} \label{dry_spot}
\end{figure}

However, the cells in ALICE-2 were not suitable to control the heat supply or to measure it. It did not permit
to change the bubble position with respect to the cell either. These limitations have been overcome in the
experimental H$_2$ setup that makes use of the magnetic levitation facility \cite{Regis,Mag} at CEA-Grenoble. A
cylindrical cell of 8~mm diameter and 5~mm height is filled with H$_2$ at critical density. The sapphire
transparent bases of the cell are good heat conductors in the cryogenic temperature range ($T_c=33$K for H$_2$).
The lateral cell wall is made of stainless steel which is on the contrary the thermal insulator. The fluid is
heated by one of the windows while the temperature of the other is maintained by the temperature regulation
system which permits us to measure the heat flux removed from the cell. The earth gravity is compensated by
magnetic forces within 2\% in the cell volume. The position of the cell with respect to the magnetic field is
chosen so that the residual force positions the bubble against the heating window. The growth of the dry spot
can thus be directly observed and is illustrated in Fig.~\ref{dry_spot}. Small vapor bubbles grow and depart
from the heater under the action of the residual magnetic forces. The bubble spreading begins however for the
unique (largest) vapor bubble. One can see that the same mechanism of the dry spot spreading works also
relatively far from the critical point. Indeed, the experiments were performed for up to 3\% deviation from
$p_c$.

\section{Conclusions}

This paper deals with the vapor recoil model which describes the triggering of the boiling crisis. The results
of both simulations and experiments in low gravity confirm this model. However, further studies are necessary.
The simulations of the bubble departure under the influence of gravity or/and externally imposed flow should be
performed to determine $q_{CHF}$, the heat flux necessary to trigger the boiling crisis. This is a complex
problem because of two reasons. First, the heat transfer singularity at the contact line needs to be resolved.
Second, the bubble surface position (including the contact line) needs to be determined rigorously, without
employing any hypotheses on the interface shape like ``liquid microlayer" hypothesis used conventionally
\cite{Dhir} to describe the contact line heat transfer.

The experiments are also needed to be carried on. The drop spreading will be studied quantitatively, both under
magnetic levitation and in microgravity on board of the International Space Station in the CNES DECLIC
apparatus.

\acknowledgements

This work was partially supported by CNES. We thank all of the ALICE team and everyone involved into the Mir
missions.

\end{document}